\documentclass[aps,showpacs,nofootinbib,subscriptaddress,twocolumn]{revtex4}

\usepackage{graphicx}
\usepackage{bm}
\usepackage{amsmath}
\usepackage{amssymb}
\usepackage{hyperref}

\newcommand{\nslash}{\kern 0.2 em n\kern -0.50em /}
\newcommand{\kslash}{\kern 0.2 em k\kern -0.45em /}
\newcommand{\pslash}{\kern 0.2 em p\kern -0.50em /}
\newcommand{\Sslash}{\kern 0.2 em S\kern -0.50em /}
\newcommand{\Pslash}{\kern 0.2 em P\kern -0.50em /}
\newcommand{\Dslash}{\kern 0.2 em D\kern -0.65em /\kern 0.15em}

\begin{document}

\title{Boer-Mulders function of the pion in the MIT bag model}

\newcommand*{\SEU}{Department of Physics, Southeast University, Nanjing
211189, China}\affiliation{\SEU}
\newcommand*{\PKU}{School of Physics and State Key Laboratory of Nuclear Physics and
Technology, \\Peking University, Beijing 100871,
China}\affiliation{\PKU}
\newcommand*{\CHEP}{Center for High Energy Physics, Peking University, Beijing 100871, China}\affiliation{\CHEP}

\author{Zhun Lu}
\affiliation{\SEU}
\author{Bo-Qiang Ma}\email{mabq@pku.edu.cn}\affiliation{\PKU}\affiliation{\CHEP}
\author{Jiacai Zhu}\affiliation{\PKU}

\begin{abstract}
We apply the MIT bag model to study the Boer-Mulders function of the pion, a $T$-odd function that describes the transverse polarization distribution of the quark inside the pion. We simulate the effect of the gauge link through the ``one-gluon-exchange" approximation. We consider both the quark helicity nonflip and double-flip contributions. The result in the MIT bag model is compared with those in the spectator models.
\end{abstract}


\maketitle


One of the main tasks in QCD and hadron physics is to understand the transverse partonic structure of hadrons, especially the nucleon and the pion.
The inclusion of the parton transverse motion introduces new types of parton structure, the so-called transverse momentum dependent (TMD) distributions, or alternatively the three-dimensional parton distribution functions in momentum space. They extend the concept of traditional Feynman distribution functions and encode a wealth of new information on the nucleon
structures~\cite{sivers,anselmino95,bhs02,collins02,belitsky,Boer:2003cm,Bacchetta:2006tn} that cannot be
described merely by the leading-twist collinear picture. Of particular interests are
the leading twist $T$-odd TMD
distribution functions, such as the Sivers function~
\cite{sivers,anselmino95} and the Boer-Mulders function~\cite{bm}.
They arise from the correlation between the nucleon/quark
transverse spin and the quark transverse momentum, and they can account for the polarized and unpolarized spin asymmetries in the
semi-inclusive deeply inelastic scattering
(SIDIS)~\cite{Airapetian:2010ds,Alekseev:2010rw,Mkrtchyan:2007sr,:2008rv,
compass2009un,Airapetian:2012yg}
and the Drell-Yan~\cite{NA10,Zhu:2006gx,Zhu:2008sj} processes.

As a spin-0 hadron, the pion has a simpler partonic structure than that of the nucleon, i.e., in leading twist there are two TMDs of the pion, the unpolarized TMD $f_1(x,\bm k_T^2)$ and the Boer-Mulders function $h_1^\perp(x,\bm k_T^2)$.
However, the pion TMDs are less known from experiments than those of the proton since they cannot be probed in the SIDIS. Theoretically, the Boer-Mulders function of the pion has been studied by lattice calculation~\cite{Brommel:2007xd} and model calculations~\cite{Lu04,Lu05,Meissner:2008ay,Gamberg:2009uk,Gamberg:2009ma,Pasquini:2012}. In the latter case, different treatments on the gauge link have been used, namely, the one-gluon exchange approximation~\cite{Lu04,Lu05,Meissner:2008ay,Pasquini:2012} and the nonperturbative eikonal methods~\cite{Gamberg:2009uk,Gamberg:2009ma}, which take into account higher
order gluonic contributions, respectively.
In this paper, we study the Boer-Mulders function of the pion using an alternative model, the MIT bag model~\cite{Chodos:1974je}. This model has been applied to study the TMDs of the proton, including the $T$-even distributions~\cite{Avakian:2010ae}, the Sivers functions~\cite{Yuan:2003wk,Cherednikov:2006zn,Courtoy:2008dn}, and the Boer-Mulders functions~\cite{Yuan:2003wk,Courtoy:2009pc}.
The calculation of the $T$-odd TMDs by the MIT bag model has produced their main features, for instance, the sign and the Burkardt sum rule~\cite{Burkardt:2004ur} for the Sivers function~\cite{Courtoy:2008dn}, and the sign for the Boer-Mulders function~\cite{Burkardt:2007xm}. Therefore, it is worthwhile to use the same model to study the Boer-Mulders function of the pion.

Unlike the Boer-Mulders function of the proton, which can be probed in both the SIDIS process and the Drell-Yan process, the Boer-Mulders function of the pion may only be detected in the Drell-Yan process.  Fortunately, the new $\pi N$ Drell-Yan program will be conducted by COMPASS~\cite{Quintans:2011zz} at CERN very soon; also there is a $\pi N$ Drell-Yan plan proposed by SPASCHARM~\cite{Abramov:2011zza}. The upcoming Drell-Yan experiments can achieve unpolarized and polarized scattering, so they will provide the opportunities~\cite{Lu:2011qp,Lu:2011pt} to access the chiral-odd TMDs of the pion as well as the nucleon.


The quark-quark correlation function for the pion has the form
\begin{equation}
\Phi_{ij}(x,\bm k_T)=\int{d\xi^- d^2\bm{\xi}_T\over (2\pi)^3}\langle P
|\psi_j(0)\mathcal{L}(0,\xi)\psi_i(\xi)|P\rangle\bigg{|}_{\xi^+=0},\label{corr}
\end{equation}
where $k^+=xP^+$, and
\begin{equation}
\mathcal{L}(0,\xi) = \mathcal{P}e^{-ig \int_0^\xi d\eta A(\eta)}
\end{equation}
is the gauge link (Wilson line)
connecting the two different space-time points $0$ and $\xi$ by all possible
ordered paths followed by the gluon field $A$
running along a process-dependent path. In this work we calculate the Boer-Mulders function in the SIDIS process.

The leading-twist TMDs of the pion can be obtained from the correlator
$\Phi(x,\bm k_T)$ by the following traces:
\begin{align}
f_1(x,\bm k_T^2) &= {1\over 2} \text{Tr}\left[\Phi(x,\bm k_T) \gamma^+\right],\\
{\epsilon^{\alpha\rho}k_{T \rho}^{}\over M}h_1^\perp(x,\bm k_T^2)&={1\over 2} \text{Tr}\left[\Phi(x,\bm k_T) i\sigma^{+\alpha}\gamma_5\right].
\end{align}

The TMD distribution $f_1(x,\bm k_T^2)$ can be calculated straightforward in the MIT bag model, in which the quark fields are expressed in the following general form:
\begin{widetext}
\begin{equation}
\Psi_\alpha(\vec x, t) = \sum_{n>0,\kappa=\pm 1, m=\pm 1/2} N
\{b_\alpha(n\kappa m)\psi_{n\kappa jm}(\vec x,t)+d^\dag_\alpha(n\kappa m)\psi_{-n-\kappa jm}(\vec x,t)\}
\end{equation}
\end{widetext}
where $\psi$ is the wavefunction in the position space.
After performing the Fourier transformation, one obtains the momentum space wavefunction of the quark~\cite{Chodos:1974je},
\begin{equation}
\varphi_m(\bm k)=i\sqrt{4\pi}NR_0^3
\left(
\begin{array}{c}
  t_0(k)\,\chi_m \\
  \bm \sigma \cdot \hat {\bm k}  \, t_1(k)\,\chi_m
  \end{array}\right), \label{wav1}
\end{equation}
and the wavefunction of the antiquark,
\begin{equation}
\phi_m(\bm k)=i\sqrt{4\pi}NR_0^3
\left(
\begin{array}{c}
 \bm \sigma \cdot \hat {\bm k} \, t_1(k)\,\chi_m  \\
 t_0(k)\,\chi_m
  \end{array}\right),\label{wav2}
\end{equation}
  where $\hat {\bm k} = {\bm k\over k}$ is a unit vector with $k=|\bm k|$, $\omega\approx 2.04$ for the lowest mode, $R_0$ the bag radius, $\chi_m$ the Pauli spinor, $\sigma$ the Pauli matrix, and $N$ the normalization factor with the form
\begin{equation}
N=\left({\omega\over 2R_0^3(\omega -1 ) j_0^2(\omega) }\right)^{1\over 2}.
\end{equation}
The functions $t_i(k)$ are calculated from
\begin{equation}
t_i(k)=\int_0^1 u^2 d uj_i(ukR_0) j_i(u\omega),
\end{equation}
where $j_i$ are the spherical Bessel functions.

Using the isospin symmetry and charge-conjugation operation, the unpolarized TMDs of the charged pion can be connected by
\begin{equation}
 f_1^{u/\pi^+} =f_1^{\bar{d}/\pi^+} =
f_1^{\bar{u}/\pi^-} =f_1^{d/\pi^-} \equiv  f_{1\pi}.
\end{equation}
The function $f_{1\pi}$ can be calculated by inserting the quark field in the MIT bag model into the correlator (\ref{corr}) in the absence of the gauge link
\begin{align}
f_{1\pi}(x,k_T) = {4\pi N^2E_\pi R^6_0 \over (2\pi)^3 }\left(t_0^2(k)+2\hat k_zt_0(k)t_1(k)+t_1^2(k)\right),
\end{align}
where $\hat k_z ={ k_z\over k}$, and $k_z=xM_\pi-\varepsilon$ with $\varepsilon=\omega/R_0$.
The distribution for the neutral pion is a half of $f_{1\pi}$.

\begin{figure}
  \includegraphics[width=\columnwidth]{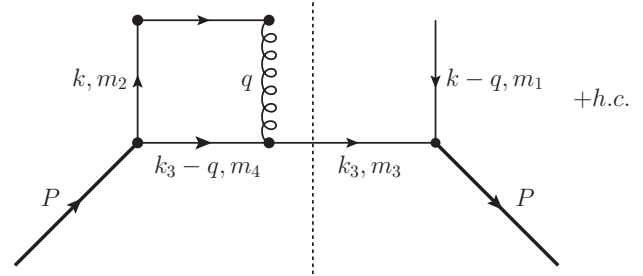}
  \caption{One gluon exchange contribution to the pion Boer-Mulders function. The graph has been drawn using JAXODRAW\cite{Binosi:2008ig}. }\label{fig1}
\end{figure}

The MIT bag model has also been extended to calculate $T$-odd TMDs~\cite{Yuan:2003wk,Courtoy:2008dn,Courtoy:2009pc}, such as the Sivers function and the Boer-Mulders function of the nucleon.
As in the original MIT bag model there is no explicit gluon degree of freedom, which is crucial for nonzero $T$-odd TMDs; in these calculations the effect of the gauge link is incorporated by introducing ``one-gluon-exchange"~\cite{Yuan:2003wk,Courtoy:2008dn,Courtoy:2009pc} or invoking instanton effects~\cite{Cherednikov:2006zn}.
In our calculation of the pion Boer-Mulders function, we follow the former approach to expand the gauge link to order $\mathcal{O}(g)$ to obtain
the expression
\begin{widetext}
\begin{align}
h_{1\pi}^{\perp}(x,\bm k_T^2) =& -2ig_s^2{M_\pi E_\pi\over k_x}
\int {d^2 \bm q_T \over (2\pi)^5}{1\over q_T^2} \sum_{m_1 m_2 m_3 m_4}
 T^a_{ij} \bar{T}^a_{kl}\langle P_\pi|b^{\dag i}_{m_1}b^{j}_{m_2}d^{\dag k}_{ m_3}d^{l}_{m_4} |P_\pi\rangle \nonumber\\
& \times \varphi_{m_1}^\dag(\bm k-\bm q_T) \gamma^0 \gamma^+\gamma^2\gamma_5
 \varphi_{m_2} (\bm k) \int {d^3 \bm k_3 \over (2\pi)^3}
 \phi_{m_3}^\dag(\bm k_3)\gamma^0 \gamma^+ \phi_{m_4}(\bm k_3 -\bm q_T),\label{h1pi}
\end{align}
\end{widetext}
where we have used the covariant gauge. We also point out that the calculated distribution is for the semi-inclusive DIS process.
The corresponding diagram is shown in Fig.~\ref{fig1}.
In Eq.(\ref{h1pi}) we use $T^a$ and ${\bar T}^a$ (the conjugate representation) to denote the Gell-Mann matrices associated with the quark and the antiquark, respectively.
They are related by
\begin{equation}
\bar{T}^a=-(T^a)^* = -(T^a)^T \label{color}.
\end{equation}
Defining
\begin{equation}
C_{m_1 m_2, m_3 m_4} = T^a_{ij} \bar{T}^a_{kl}\langle P_\pi|b^{\dag i}_{m_1}b^{j}_{m_2}d^{\dag k}_{ m_3}d^{l}_{m_4} |P_\pi\rangle,
\end{equation}
we obtain the following nonzero spin coefficients:
\begin{equation}
C_{+ +, - -} = C_{-- , + + } =-{2\over 3},~~C_{+ -, - +} = C_{-+, +- } ={2\over 3}.\label{spincoe}
\end{equation}
These coefficients  have already included the contribution from the color factor
\begin{equation}
{1\over 3}\sum_{ijkl} \delta_{ik} \delta_{jl} (\,T^a_{ij} \bar{T}^a_{kl})
 =-{4\over 3},
\end{equation}
calculated from the color(-singlet) structure of the pion and the Gell-Mann matrices.

Inserting the bag wavefunctions (\ref{wav1}) and (\ref{wav2}), and the spin coefficients (\ref{spincoe}) into (\ref{h1pi}), we arrive at the final expression of the pion Boer-Mulders function
\begin{widetext}
\begin{equation}
h_{1\pi}^\perp(x,\bm k_T^2) =- g_s^2 {M_\pi^2\over k_x} \int {d^2q_T\over (2\pi)^2} {C^2\over q_T^2} \int{d^3 \bm k_3}(R_{++}\,R^\prime_{++} +I_{++}\,I^\prime_{++} + R_{+-}\,R^\prime_{+-} +I_{+-}\,I^\prime_{+-}),\label{pibm}
\end{equation}
\end{widetext}
where
 \begin{equation}
 C={2\over\sqrt{3}}  {4\pi N^2 R_0^6 \over (2\pi)^3}.
 \end{equation}
In Eq.~(\ref{pibm}) 
the real functions $R_{m_1m_2}$, $I_{m_1 m_2}$, $R_{m_3m_4}^\prime$, and $I_{m_3 m_4}^\prime$ are defined as
\begin{align}
&\varphi_{m_1}^\dag \gamma^0 \gamma^+\gamma^2\gamma_5
 \varphi_{m_2} \nonumber\\
=& \frac{i}{\sqrt{2}}4\pi N^2 R_0^6(R_{m_1m_2}+iI_{m_1m_2}),
\end{align}
\begin{align}
&\phi_{m_3}^\dag\gamma^0 \gamma^+ \phi_{m_4}\nonumber\\
=&\frac{1}{\sqrt{2}} 4\pi N^2 R_0^6(R_{m_3m_4}^\prime+iI_{m_3m_4}^\prime).
\end{align}

There are totally 16 functions, among which 8 are independent and have the following forms:
\begin{widetext}
\begin{align}
R_{++} & = \hat k_x t_0(k^\prime) t_1(k) -\hat k^\prime_x t_0(k) t_1(k^\prime)
+( \hat k_z^\prime \hat k_x
- \hat k_x^\prime\hat k_z )t_1(k^\prime) t_1(k) ,\label{rpp}\\
R_{+-} & = -t_0(k^\prime) t_0(k) - \hat k_z t_0(k^\prime) t_1(k) -\hat k^\prime_z t_0(k) t_1(k^\prime)
+( -\hat k_z^\prime \hat k_z
- \hat k_x^\prime\hat k_x+\hat k_y^\prime\hat k_y )t_1(k^\prime) t_1(k), \\
I_{++} & = \hat k_y t_0(k^\prime) t_1(k) +\hat k^\prime_y t_0(k) t_1(k^\prime)
+( \hat k_z^\prime \hat k_y
+ \hat k_y^\prime\hat k_z )t_1(k^\prime) t_1(k) ,\\
I_{+-} & =(
\hat k_x^\prime\hat k_y+\hat k_y^\prime\hat k_x )t_1(k^\prime) t_1(k),\\
R_{++}^\prime & = \hat k_3 \cdot\hat k_3^\prime t_1(k_3) t_1(k_3^\prime)
 +\hat k_{3z} t_1(k_3) t_0(k_3^\prime)
+\hat k_{3z}^\prime t_0(k_3) t_1(k_3^\prime)
+t_0(k_3)  t_0(k_3^\prime),  \\
R_{+-}^\prime & =-(\hat k_{3x}\hat k_{3z}^\prime- \hat k_{3z}\hat k_{3x}^\prime)t_1(k_3) t_1(k_3^\prime)
 -\hat k_{3x} t_1(k_3) t_0(k_3^\prime)
+\hat k_{3x}^\prime t_0(k_3) t_1(k_3^\prime),\\
I_{++}^\prime & = (\hat k_{3x}\hat k_{3y}^\prime- \hat k_{3y}\hat k_{3x}^\prime)t_1(k_3) t_1(k_3^\prime), \\
I_{+-}^\prime & =(\hat k_{3y}\hat k_{3z}^\prime- \hat k_{3z}\hat k_{3y}^\prime)t_1(k_3) t_1(k_3^\prime)
 +\hat k_{3y} t_1(k_3) t_0(k_3^\prime)
-\hat k_{3y}^\prime t_0(k_3) t_1(k_3^\prime) ,\label{ipm}
\end{align}
\end{widetext}
where $k^\prime = |\bm k -\bm q_T|$, $\hat {\bm k}^\prime= {\bm k-\bm q_T\over k^\prime}$,  and
$k^\prime_3 = |\bm k_3 -\bm q_T|$, $\hat {\bm k}^\prime_3 = {\bm k_3 -\bm q_T\over k_3^\prime}$. The functions listed in Eqs.(\ref{rpp})-(\ref{ipm}) agree with the functions $I_{1,2}$,  $F_{1,2}$, $H_{1,2}$, and $J_{1,2}$ listed in the appendix of Ref.~\cite{Courtoy:2009pc}. The first two terms on the right-hand side of (\ref{pibm}) are the quark helicity nonflip contributions, while the last two terms are the contributions received from the helicity double-flip of quarks. An important observation in the bag model calculation of the proton Sivers function~\cite{Courtoy:2008dn} and Boer-Mulders function~\cite{Courtoy:2009pc} is that, apart from the  helicity nonflip contributions, the double-flip terms (especially the $R_{+-}R_{+-}^\prime$ term ) is significant and should not be ignored. In the light of this finding, in this work we consider both the helicity nonflip and double-flip contributions to the pion Boer-Mulders function.


To give a numerical estimate of the pion Boer-Mulders function in the MIT bag model, we need to fix the parameters in the model, especially the bag radius $R_0$ of the pion. In the calculation of the proton TMDs~\cite{Yuan:2003wk,Courtoy:2009pc,Courtoy:2008dn,Avakian:2010} the bag radius is determined by the relation~\cite{Chodos:1974pn}
\begin{equation}
R_0={4\over 3} {n \omega\over M_n},
\end{equation}
where $n$ is the quark (antiquark) number in the bag. Here we use the same ansatz for the bag radius of the meson. For the strong coupling $\alpha_s$,  we follow the choice $\alpha_s/(4\pi)=0.13$ in \cite{Courtoy:2009pc}, where the same model has been used to calculate the proton Boer-Mulders function. To get the appropriate tendency of the distribution at the region $x\rightarrow 1$,  we use the constraint $\delta(1-x-x_3)$ when performing the integration in (\ref{h1pi}), where $x_3 = k_3^+/P^+$.

\begin{figure*}
\includegraphics[width=0.49\textwidth]{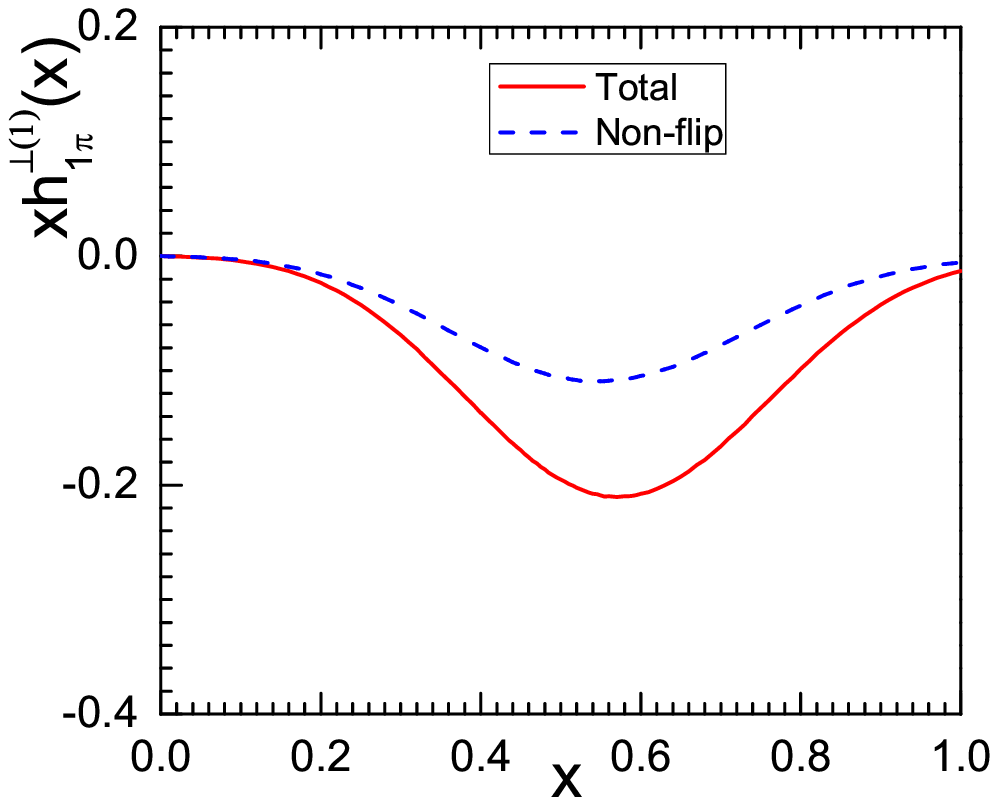}
\includegraphics[width=0.49\textwidth]{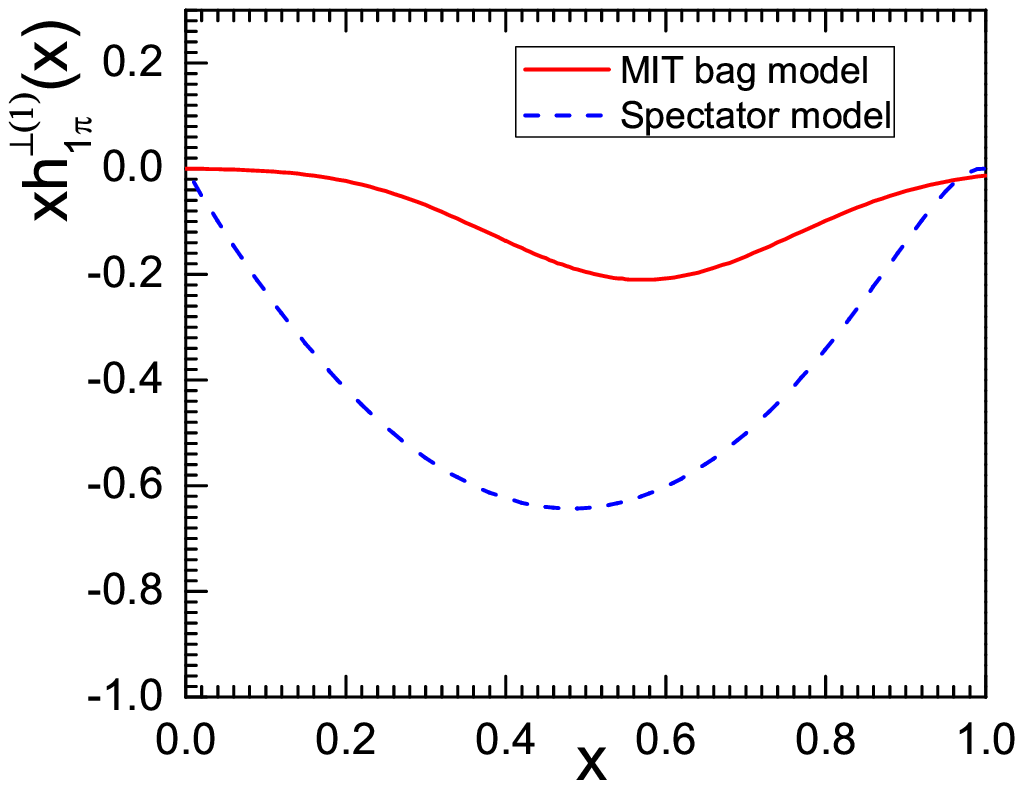}
  \caption{Left panel: The first $k_T^2$ moment of the pion Boer-Mulders function  in MIT bag model. The solid line corresponds to the full result including both the helicity nonflip and double-flip contributions. The dashed line corresponds to the result with only the helicity nonflip contribution. Right panel: Comparison of $xh^{\perp(1)}_{1\pi}(x)$ in the MIT bag model (solid line) and the spectator model (dashed line).}\label{fig2}
\end{figure*}

The left panel of Fig.~\ref{fig2} shows the first $k_T^2$ moment of the Boer-Mulders function $h_{1\pi}^{\perp(1)}(x)$, which is defined as
\begin{equation}
h_{1\pi}^{\perp (1)}(x) = \int d^2 \bm k_T \Big({\bm k_T^2\over 2 M^2}\Big) h_{1\pi}^\perp (x,\bm k_T^2).
\end{equation}
The solid and the dashed curves represent the total result and the result contributed by the quark helicity nonflip terms. The comparison of these two curves indicates that the helicity nonflip and double-flip contributions are equally important to the pion Boer-Mulders function. Our results show that $h_{1\pi}^{\perp}$ is
negative, in agreement with spectator model and lattice calculations. The sign of the pion Boer-Mulders function is also consistent~\cite{Burkardt:2007xm} with the sign of the Boer-Mulders functions of the nucleon in the MIT bag model.

In the right panel of Fig.~\ref{fig2} we compare
$xh^{{\perp}(1)}_{1\pi}(x)$ in the MIT bag model (shown by the solid
line) with that in the spectator model (shown by the dashed line)
\cite{Lu04}, where the one-gluon exchange approximation is also
used. When obtaining the two curves in the right panel of
Fig.~\ref{fig2} we use the same strong coupling
$\alpha_s/(4\pi)=0.13$ for comparison. The size of
$xh^{{\perp}(1)}_{1\pi}(x)$ in the MIT bag model is smaller than
that in the spectator model, while the $x$ dependence of the
distribution are similar in both models; that is, they peak at the
region $x\sim 0.5$.

We also point out that the size of $xh^{{\perp}(1)}_{1\pi}(x)$
in our calculation is comparable with the spectator model
calculation that employs the nonperturbative eikonal
methods~\cite{Gamberg:2009uk,Gamberg:2009ma}. The $x$ dependence of
$xh^{{\perp}(1)}_{1\pi}(x)$  in these two different calculations
differ from each other since in \cite{Gamberg:2009uk,Gamberg:2009ma}
the distribution peaks at $x\sim 0.2$.

Similar to the MIT bag model calculations for the proton TMDs, our
calculations are performed at the low energy bag scale, and we have
not considered the evolution effect during the entire calculation.
Recently substantial progress~\cite{Aybat:2011zv,Aybat:2011ge} on
the evolution of the proton TMDs has been achieved. A potential
issue for future study is to investigate if the same approach can be
applied to the pion TMDs, especially the Boer-Mulders function,
where the MIT bag model calculations could be the initial inputs at
the low energy scale. Only with full knowledge of the initial inputs
and the evolution of the pion TMDs, we can get more precise
predictions of the experiments.


In summary, we have applied the MIT bag model to study the TMDs of
the pion. Particularly, we calculated the pion Boer-Mulders
function, which is a $T$-odd chiral-odd distribution. To
obtain a nonzero result, the effect of the gauge link is simulated by
introducing the ``one-gluon-exchange" effect. We consider both the
helicity nonflip and double-flip contributions to the pion
Boer-Mulders function. We estimated the pion Boer-Mulders function
numerically, showing that it is negative in the MIT bag model. We
compare our result with the available  spectator model calculations.
Our study provides further knowledge on the transverse parton
structure of the pion.

This work is partially supported by National Natural Science
Foundation of China (Grants No.~10905059, No.~11005018,
No.~11021092, No.~10975003, No.~11035003, and No.~11120101004),
   by SRF for ROCS, SEM, and by the Teaching and Research Foundation for
Outstanding Young Faculty of Southeast University.

\end{document}